\documentclass[%
 reprint,
superscriptaddress,
 amsmath,amssymb,
 aps,
 pra,
]{revtex4-2}

\usepackage{graphicx}
\usepackage{dcolumn}
\usepackage{bm}
\usepackage{hyperref}
\usepackage[mathlines]{lineno}
\usepackage{lipsum}
\usepackage{xcolor}
\usepackage{siunitx}
\usepackage{braket} 
\usepackage{placeins} 
\usepackage{comment}

\usepackage{tocloft}
\usepackage{array}
\newcolumntype{R}[1]{>{\raggedleft\arraybackslash}p{#1}}

\begin{document}


\preprint{APS/123-QED} 

\title{Atom and spin resolved imaging in a single shot}

\author{Tobias~Hammel}
\thanks{Corresponding author: hammel@physi.uni-heidelberg.de}

\author{Maximilian~Kaiser}
\affiliation{Physikalisches Institut der Universität Heidelberg, Im Neuenheimer Feld 226, 69120 Heidelberg, Germany}

\author{Daniel~Dux}
\affiliation{Physikalisches Institut der Universität Heidelberg, Im Neuenheimer Feld 226, 69120 Heidelberg, Germany}

\author{Matthias~Weidemüller}
\affiliation{Physikalisches Institut der Universität Heidelberg, Im Neuenheimer Feld 226, 69120 Heidelberg, Germany}

\author{Selim~Jochim}
\affiliation{Physikalisches Institut der Universität Heidelberg, Im Neuenheimer Feld 226, 69120 Heidelberg, Germany}

\date{\today}


\begin{abstract}
    We report on an imaging scheme for quantum gases that enables simultaneous detection of two spin states with single-atom resolution. It utilizes the polarization of the emitted photons during fluorescence by choosing appropriate internal states of lithium-6 atoms in a magnetic field. This scheme can readily be implemented to obtain in-situ spin correlations in a wide variety of experimental settings.

\end{abstract}

\maketitle

High-resolution fluorescence imaging is a broadly applicable technique for probing the fundamental properties of cold atom systems. Among different fluorescence detection methods in quantum gas experiments, there has been significant progress enabling, for example, single-atom resolved imaging in quantum gas microscopes \cite{Gross2021} and in free space \cite{Bergschneider_2018, B_cker_2009}. In such experimental setups, the detection of the internal state of the atoms is a technical challenge. These states can be represented as pseudo-spins and are used to realize, e.g., multi-component Fermi gases. Hence, their resolution is essential for gaining insight into the fundamental correlations of the system.

One way to obtain spin-resolved information is to project the spin states onto other, more easily detectable, properties. A prominent example is the Stern-Gerlach effect \cite{Gerlach1922}, mapping spin states to spatial positions. This has been utilized to achieve spin resolution in a variety of ways in quantum gas experiments, including time-of-flight evolution for continuous systems \cite{StamperKurn, Schmaljohann2004, Taie2010, fallani2023} or using in- and out-of-plane superlattices \cite{Boll2016, Koepsell2020, Preiss2015} in quantum gas microscopes.

An alternative method, realized in \cite{Bergschneider_2018}, is the imaging of two spins in quick succession. There, the spin information is encoded in the spectral response to the imaging pulses, where the degeneracy of the energy of different hyperfine states is lifted by the Zeeman effect in an external magnetic field, thereby making the states individually addressable.

In this paper, we present an imaging scheme that assigns the spin to the individual atoms in a single measurement, without the use of complex auxiliary optics. We utilize the polarization of the emitted fluorescence light to infer the spin state of the atoms in a polarization-sensitive imaging setup. With this approach, spin and single-atom resolved images can be obtained efficiently in quantum gas experiments.\\ 

\begin{figure*}[ht]
    \centering
    \includegraphics[width=1.0\linewidth]{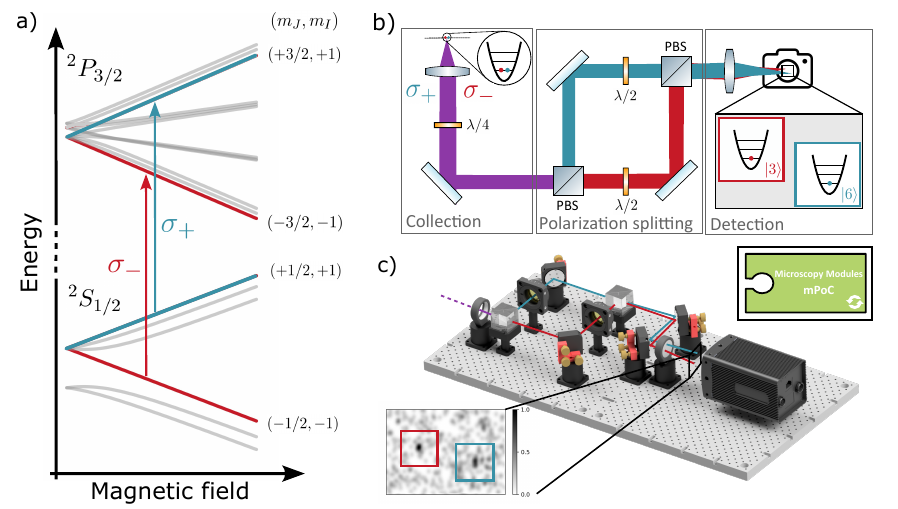}
    \caption{\textbf{Spin and atom resolved imaging scheme} a) Level structure of the $2s_{1/2}$ ground and the $2p_{3/2}$  excited state in $^6\text{Li}$ and their magnetic field dependence. The closed cycling transitions for the stretched states $\ket{3}=\ket{m_J=-1/2, m_I=-1}$ and $\ket{6}=\ket{m_J=+1/2, m_I=+1}$ are marked in red and blue, and the corresponding polarization is assigned. b) Schematic representation of the optical setup enabling simultaneous imaging of both spin states. In the collection part, the fluorescence light from an arbitrary system, here depicted as the harmonic oscillator ground state of two fermions with different spins, is collected by a lens. The overlapping left- and right-handed fluorescence light emitted from the $\sigma_+$ and $\sigma_-$ transitions is transformed into orthogonal linear polarizations by a quarter-waveplate. In the polarization-splitting section, PBS cubes are used to split the polarizations and then recombine them at slightly different angles. The half-waveplates in each path are used such that each beam is transmitted exactly once, which optimizes the extinction ratio given by the cubes and hence minimizes cross-coupling between the two images. In the detection part, the two beams are focused onto two parts of the same camera chip, where on the left (right) hand side the signal of atoms in state $\ket{3}$ ($\ket{6}$) is located. c) A rendering of the modular imaging setup used in this experiment. Technical details of the implementation into the \textit{Heidelberg Quantum Architecture} are presented in \cite{Hammel2025} and referenced by the microscopy module puzzle piece. In the inset, an experimental image of the system depicted in b) is presented, namely, the ground state of a harmonic oscillator filled with two fermions of different spins. The image is post-processed as described in \cite{Bergschneider_2018}: First, the image is binarized by setting each pixel to 0 or 1 based on the raw pixel value relative to a threshold. Then, the data are folded with a Gaussian to generate a low-pass image, which is subsequently analyzed using a peak-finding routine.}
    \label{fig:scheme}
\end{figure*}

The basic idea of this scheme is shown in Figure~\ref{fig:scheme}, utilizing ideas of free-space fluorescence imaging presented in \cite{Bergschneider_2018}. There, atoms are brought to fluorescence by resonant laser light. Closed transitions are particularly useful because spontaneous decay after excitation repopulates only the initial state, thereby maintaining a closed optical cycle. In Figure~\ref{fig:scheme}a) the closed transitions in the case of $^6\text{Li}$ are marked in blue and red. At high magnetic fields, the Zeeman splitting is large enough to address them spectroscopically. Previously, spin resolution has been achieved by taking two consecutive images at different frequencies of the fluorescence light, one for each spin state \cite{Bergschneider_2018}.

In Figure~\ref{fig:scheme}b), we show the schematic of our setup that can image both spin states simultaneously. The central point of this new scheme is that the two spin states couple to closed optical transitions, emitting circularly polarized light of opposite handedness, which enables spatial separation of their fluorescence light. During fluorescence imaging, the left- and right-handed, circularly polarized light emitted from the $\sigma_{+}$ and $\sigma_{-}$ transitions is collected using a high-NA objective. A quarter-waveplate transforms the circular polarizations into orthogonal linear polarizations. Then, the polarizations are split into two paths using a polarizing beam splitter cube (PBS), with each light field carrying information about one of the two spin states. Both light fields are then recombined on a second PBS, with the additional degree of freedom to adjust their relative orientation. Hence, they can be focused onto an imaging sensor in two different locations, one for each polarization. This results in two simultaneous images of the same spatial region, one per spin, on the same camera chip. The inset of Figure~\ref{fig:scheme}b) shows the schematic of such an image, where on the left side of the chip, the information of atoms in one state is collected, and on the right side, for atoms in another state. 

To demonstrate this scheme, we prepare two $^6\text{Li}$ atoms of different spins (namely $\ket{1}$ and $\ket{3}$) in the ground state of an optical tweezer and image them spin and atom resolved in a single picture. For an optical tweezer array, this scheme is implemented in \cite{Jain2025}. In Figure~\ref{fig:scheme}c), a rendering of our optical imaging setup is shown, which is connected to our experiment in a modular way described in \cite{Hammel2025} and named \textit{Heidelberg Quantum Architecture}. The optics modules, or ``pieces of cake", offer several advantages, including convenient exchange and implementation of imaging modules with different magnifications and cameras, and the disentanglement of the imaging setup from the modules that generate the optical potentials. 

In the inset of Figure~\ref{fig:scheme}c) an image of the two fermion system is shown, where the left signal stems from the atom in state $\ket{3}$, while the right signal comes from the atom, which was transferred from the $\ket{1}$ state to the $\ket{6}$ state before imaging using a microwave pulse. In particular, the atoms occupy the same spatial position in the optical tweezer but are clearly separated on the camera chip because of their different paths through the polarization-splitting setup.

This imaging scheme allows for simultaneous imaging of two spin states. The characteristic feature is the outstanding simplicity of the optical setup, which can be readily implemented in a variety of existing experiments, ranging from continuous fermionic and spinful bosonic systems to quantum gas microscopes. A high-fidelity implementation and quantitative analysis of this scheme is presented in \cite{Jain2025}. Furthermore, it relaxes technical restrictions on cameras and on the optical potentials and schemes used to identify the spins. Hence, it can enable the use of advanced techniques for high-resolution imaging of quantum gases \cite{Asteria_2021, brandstetter2024, Su2025} in experiments, where technical restrictions previously proved prohibitive.

\begin{acknowledgments}

This work has been supported by the Heidelberg Center for Quantum Dynamics, the DFG Collaborative Research Centre SFB 1225 (ISOQUANT), Germany’s Excellence Strategy EXC2181/1-390900948 (Heidelberg Excellence Cluster STRUCTURES), and the European Union’s Horizon Europe programme HORIZON-CL4-2022-QUANTUM-02-SGA via the project 101113690 (PASQuanS2.1).

\textit{Data Availability} The data that support the findings of this article are openly available \cite{Data}.

\textit{Competing Interest} The authors declare no competing interests.

\textit{Correspondence and requests for materials} should be addressed to T.H. (hammel@physi.uni-heidelberg.de).

\end{acknowledgments}
\newpage
\bibliography{Main}

\end{document}